\documentclass[aps,twocolumn,superscriptaddress,nofootinbib]{revtex4}

\usepackage{latexsym}
\usepackage{hyperref}
\usepackage{epsfig}
\usepackage{graphics}
\usepackage{amsmath}
\usepackage{xspace}
\usepackage{amssymb}
\usepackage{amsthm}

\def\tr{\mbox{tr}}
\theoremstyle{plain}
\newtheorem{theorem}{Theorem}
\newtheorem{lemma}{Lemma}

\newtheorem{proposition}{Proposition}
\theoremstyle{definition}

\theoremstyle{remark}

\begin{document}

\title{Entanglement, correlations, and the energy gap in 
many-body quantum systems}

\author{Henry~L.~Haselgrove} %
\email{hlh@physics.uq.edu.au}  
\affiliation{School of Physical
Sciences, The University of Queensland, Brisbane 4072, Australia}
\affiliation{Institute for Quantum Information, California
Institute of Technology, Pasadena CA 91125, USA}
\affiliation{Information Sciences Laboratory, Defence Science and
Technology Organisation, Edinburgh 5111 Australia}
\author{Michael~A.~Nielsen} 
\email{nielsen@physics.uq.edu.au}
\homepage[\\ URL: ]{http://www.qinfo.org/people/nielsen/}
\affiliation{School of Physical
Sciences, The University of Queensland, Brisbane 4072, Australia}
\affiliation{School of Information Technology and Electrical Engineering,
The University of Queensland, Brisbane 4072, Australia}
\affiliation{Institute for 
Quantum Information, California Institute of Technology, Pasadena
CA 91125, USA}
\author{Tobias~J.~Osborne}
\email{T.J.Osborne@bristol.ac.uk} 
\affiliation{School of Mathematics,
University of Bristol, University Walk, Bristol BS8 1TW, United
Kingdom}
\date{\today}

\begin{abstract}
  What correlations are present in the ground state of a many-body
  Hamiltonian?  We study the relationship between ground-state
  correlations, especially entanglement, and the \emph{energy gap}
  between the ground and first excited states.  We prove several
  general inequalities which show quantitatively that ground-state
  correlations between systems not directly coupled by the Hamiltonian
  necessarily imply a small energy gap.
\end{abstract}
\maketitle

\section{Introduction}

%
%
A central problem in physics is characterizing the ground state of a
many-body Hamiltonian.  Of particular interest is the problem of
understanding the correlations in the ground state of such systems.
As an outgrowth of that interest, there has recently been considerable
work on understanding the \emph{non-classical} correlations in the
ground state, that is, the \emph{ground-state entanglement}.  Some
recent work on this problem, with further references,
includes~\cite{Haselgrove03b,Vidal03b,Latorre03a,Tessier03a,Costi03a,Hines03a,Osterloh02a,Osborne02a,Scheel02a,Wang02b,Oconnor01a,Gunlycke01a,Nielsen98d}
(c.f. also~\cite{Aharonov03b,Kempe03a,Aharonov02a,Farhi01a}, and
references therein).

%
%
The purpose of the present paper is to prove some general inequalities
relating the ground-state correlations and entanglement to the
spectrum of the system Hamiltonian.  We will prove that the existence
of long-range correlations in the ground state implies a small energy
gap between the ground and first excited states of the system.  Our
use of ``long-range'' here is a convenient euphemism; we mean simply
correlations between subsystems not directly coupled by the system
Hamiltonian.

%
%
To be more concrete, let us describe a specific example of our
results.  Suppose we have a three-body system, with the bodies labeled
$1, 2$ and $3$.  We suppose systems $1$ and $2$ are coupled, and
systems $2$ and $3$ are coupled.  Importantly, systems $1$ and $3$ are
\emph{not} directly coupled.  This is the only assumption we make
about the system Hamiltonian.  Suppose $\psi$ is some joint pure state
of the three systems, possessing ``sufficient correlations'' between
systems $1$ and $3$, in a sense to be made precise later.  Our goal is
to relate the energy gap to the overlap $F = |\langle
\psi|E_0\rangle|$ between $\psi$ and the ground state $|E_0\rangle$ of
the system.  We will prove that:
\begin{eqnarray} \label{eq:intro-example}
  \frac{\Delta E}{E_{\rm tot}} \leq 2(1-F^2),
\end{eqnarray}
where $\Delta E$ is the energy gap, and $E_{\rm tot}$ is the total
energy scale for the system, i.e., the difference between the maximal
and minimal energies of the Hamiltonian.  The ratio of the gap to
total energy scale is an appropriate dimensionless parameter for
deciding whether a gap is small or large.  Note that rescaling of the
Hamiltonian corresponds physically just to a rescaling of time, so one
can only expect results in terms of such a dimensionless parameter; it
does not make sense to say that a gap is ``small'' in any absolute
sense --- one needs to compare it to another relevant energy scale.
The inequality~(\ref{eq:intro-example}) tells us that as the overlap
$F$ tends to one, the gap size must vanish, compared to the total
energy scale in the system, whenever the state $\psi$ exhibits
sufficiently strong correlations between systems $1$ and $3$.

%
%
Eq.~(\ref{eq:intro-example}) is just one example of the sort of
relation we'll prove.  We'll prove a variety of similar relations, for
different situations.  In particular, we'll analyze more general
coupling schemes, and will consider the relationship of correlations
to the energies of low-lying states other than the first excited state.

%
%
Our investigations may be placed in several different contexts,
including the theory of quantum phase transitions, results from
quantum many-body physics such as the Goldstone theorem, and the
theory of entanglement developed within the burgeoning field of
quantum information science.  We now briefly review these connections.

%
%
A quantum phase transition~\cite{Sachdev99a,Sondhi97a} is a
qualitative change in the properties of the ground state of a
Hamiltonian $H(g)$ as a parameter $g$ in the Hamiltonian is varied
through a \emph{critical point}, $g_c$.  The parameter $g$ might, for
example, be the value of an external magnetic field applied to a
system of spins.  Near a critical point, a system undergoing a
second-order quantum phase transition usually exhibits two related
phenomena.  The first phenomenon is truly long-range correlations in
the ground state, in the sense of correlations that decay only slowly
with distance.  The second phenomenon is a vanishing energy gap,
$\Delta E \rightarrow 0$.  These phenomena are expressed via the
relations
\begin{eqnarray}
  \frac{1}{\xi} & \propto & |g-g_c|^{\eta}, \,\,\,\,  
  \Delta E \propto |g-g_c|^{z \eta}.
\end{eqnarray}
In these relations, $\xi$ is the characteristic length scale on which
correlations occur in the system, $\Delta E$ is the energy gap, as
before, and $\eta$ and $z$ are constants known as \emph{critical
  exponents}.  Remarkably, the exact values of $\eta$ and $z$ do not
depend on the particular details of the microscopic interactions in
the system, but only on a small number of system parameters, such as
dimensionality and symmetry; this phenomenon is known as universality.
The exponent $z$ is known as the \emph{dynamical critical exponent},
and relates the way in which the energy gap vanishes to the way
long-range correlations emerge near the critical point.  In
particular, we see that $\Delta E \propto \xi^{-z}$, so that provided
the critical exponent $z$ is positive, the energy gap and the
correlations length behave inversely to one another; as the gap
becomes small, the correlation length becomes large, and vice versa.

%
%
Clearly, the study of the dynamical critical exponent $z$ has much in
common with the questions we are pursuing here.  However, there are
many significant differences.  In particular, work on quantum phase
transitions usually requires working in the thermodynamic limit of an
infinite number of systems, and often requires additional symmetry
assumptions, such as translational invariance.  Although numerous
physical examples have suggested that it is generally true that
correlations decay exponentially with the size of the energy gap, it
is only relatively recently that a general proof of this fact has been
provided~\cite{Hastings03a}, for systems in the thermodynamic limit.

%
%
%
In contrast, our results apply for any many-body quantum system,
whether in the thermodynamic limit or not, and do not require any
additional symmetry assumptions, such as translational invariance.
Thus, our results complement those obtained in the study of quantum
phase transitions.

%
%
Another context for our work is a classic result from quantum
many-body physics, the non-relativistic Goldstone
theorem~\cite{Lange66a,Goldstone61a,Goldstone62a} (see Chap.~9
of~\cite{Auerbach94a} for a review), which shows that diverging
correlations imply a vanishing energy gap.  However, as with the case
of work on quantum phase transitions, these results are complementary
to ours, in that they rely on having infinite systems, and typically
require additional symmetry assumptions.

%
%
An intriguing aspect of our results is that they make considerable use
of techniques developed in the new field of quantum information
science\footnote{See~\cite{Nielsen00a,Preskill98c} for reviews and
  further references.}, especially techniques developed for the study
of entanglement.  Thus, our paper illustrates a general idea discussed
elsewhere~\cite{Vidal03b,Osborne02b,Nielsen02e,Preskill00a,Nielsen98d}
(c.f. also~\cite{Aharonov99b}), namely, that quantum information
science may provide tools and perspectives for understanding the
properties of complex quantum systems, complementary to the existing
tools used in quantum many-body physics.

%
%
We begin the paper in Sec.~\ref{sec:toy-model} with a simple,
easily-understood toy model that illustrates many of the main physical
ideas of the paper in a heuristic way.  Much of the remainder of the
paper is devoted to generalizations and formalization of the ideas in
Sec.~\ref{sec:toy-model}.  Interestingly, the mathematics that arises
when generalizing and formalizing the results of
Sec.~\ref{sec:toy-model} leads in a natural way to other problems of
great physical interest, and exploring these connections is a theme of
the paper.

%
%
The next section of the paper, Sec.~\ref{sec:framework}, sets up a
general framework for our investigations, introducing a convenient
language to describe complex interactions involving many bodies, and
precisely framing the questions we address in this language.
Sec.~\ref{sec:unification} is the core of the paper, presenting a
series of general results connecting long-range ground-state
correlations to the energy gap and other properties of the low-lying
states.  Sec.~\ref{sec:qecc} explores an intriguing connection of our
results to the theory of quantum error-correcting codes.  Finally,
Sec.~\ref{sec:discussion} concludes with a discussion of open
questions.

\section{Invitation: a toy model}
\label{sec:toy-model}

%
%
We begin with a toy model which illustrates in a simple setting many
of the important physical ideas developed in more detail later in the
paper.  Our purpose in presenting these ideas first in a simple form
is to keep the underlying physical ideas distinct from some of the
mathematical complexities of later sections.  Keep in mind, however,
that some of these later mathematical complexities reveal surprising
connections to other physical problems whose importance may not be
apparent in the simplified setting discussed in this section.

%
%
Our toy model is a system of three qubits (spin-$\frac 12$ systems)
arranged in a line.  We label the qubits $1$, $2$, and $3$.  Suppose
the qubits are coupled by a Hamiltonian, $H$, which contains only
nearest-neighbour interactions, and so can be written $H = H_{12} +
H_{23}$.  Note that single-qubit contributions to the Hamiltonian can
be included in the interactions $H_{12}$ and $H_{23}$.  For our
purposes all that matters is that there are no couplings between
qubits $1$ and $3$.  Suppose the ground state of $H$, $|E_0\rangle$,
is non-degenerate, with corresponding ground state energy $E_0$.
Suppose the gap to the energy of the first excited state is $\Delta
E$.

%
%
How entangled are qubits $1$ and $3$ in the ground state,
$|E_0\rangle$?  We will prove that in order for qubits $1$ and $3$ to
approach maximal entanglement, the gap $\Delta E$ must approach zero.
We will only give a heuristic argument for now, with general proofs to
follow later.  Note, in particular, that while the following argument
applies for maximal entanglement between qubits $1$ and $3$, the
results of subsequent sections can be applied to more general types of
correlation.

%
%
We begin by observing that since qubits $1$ and $3$ are nearly
maximally entangled, then
\begin{eqnarray} \label{eq:step-1}
  |E_0\rangle \approx |\psi\rangle \equiv |{\rm ME}\rangle_{13} |\phi\rangle_2,
\end{eqnarray}
where $|{\rm ME}\rangle$ is some maximally entangled two-qubit state,
$|\phi\rangle$ is a single-qubit state, and subscripts indicate which
systems the states are associated with.  But since $|E_0\rangle
\approx |\psi\rangle$, the expectation energy for $|\psi\rangle$ must
also be close to $E_0$,
\begin{eqnarray} \label{eq:step-2}
   \langle \psi| H |\psi\rangle \approx E_0.
 \end{eqnarray}
 Next, let $|{\rm ME}'\rangle$ be a two-qubit maximally entangled
 state orthogonal to $|{\rm ME}\rangle$, and define
 $|\psi_{\perp}\rangle \equiv |{\rm ME}'\rangle_{13} |\phi\rangle_2$.
 Note that $|\psi_{\perp}\rangle$ is orthogonal to $|\psi\rangle$, and
 in view of Eq.~(\ref{eq:step-1}) it must be true that
 $|\psi_{\perp}\rangle$ is approximately orthogonal to $|E_0\rangle$.
 It follows that $|\psi_{\perp}\rangle$ can be expressed,
 approximately, as a superposition of states with energies $E_1$ and
 higher, where $E_1$ is the energy of the first excited state.
 Therefore the expectation energy for $|\psi_{\perp}\rangle$ must be
 at least $E_1$:
\begin{eqnarray} \label{eq:step-3}
   \langle \psi_\perp|H|\psi_\perp \rangle \geq E_1 + \mbox{small 
     corrections}.
\end{eqnarray}
These small corrections can, in principle, be negative, and we will
see that this \emph{must} be the case, in order to be consistent with
the reasoning below.

%
%
Next, observe that the expectation energies for $|\psi\rangle$ and
$|\psi_\perp\rangle$ are the same,
\begin{eqnarray} \label{eq:step-4}
  \langle \psi|H |\psi \rangle = \langle \psi_\perp |H |\psi_\perp \rangle.
\end{eqnarray}
To see this, observe that $\langle \psi|H_{12}|\psi\rangle = \langle
\psi_\perp |H_{12}|\psi_\perp \rangle$, since the reduced density
matrices for $\psi$ and $\psi_\perp$ are \emph{identical} on the
system $12$.  A similar argument shows that the contribution to the
expectation energy from $H_{23}$ is the same from both $\psi$ and
$\psi_\perp$. Combining these results gives Eq.~(\ref{eq:step-4}).

%
%
To complete the argument, observe that
Eqs.~(\ref{eq:step-2})-(\ref{eq:step-4}) can only be consistent if
$E_0 \approx E_1 + \mbox{small corrections}$, and thus the energy gap
must itself be small.

%
%
Summarizing, the presence of nearly maximal ground-state entanglement
between sites which do not directly interact allows us to construct a
state which (a) is almost orthogonal to the ground state, and thus
must have energy of about $E_1$ or higher; but (b) looks locally very
much like the ground state, and thus must have energy approximately
$E_0$.  The only way these two facts can simultaneously be true is if
the energy gap is comparable in size to the corrections used in our
approximations.  Making this argument precise, and generalizing it
further, is the subject of subsequent sections.

\section{Framework}
\label{sec:framework}

%
%
This section introduces a framework for generalizing and formalizing
the ideas of the previous section.  We first introduce some general
language for describing interactions in many-body quantum systems,
then use this language to precisely state the main questions addressed
through the remainder of the paper.  We conclude with an overview of
our answers to these questions.

%
%
In the previous section we considered three interacting qubits, with
the restriction that the first and third qubits do not interact.  It
is helpful to introduce some language to describe more general
interactions.

%
%
Suppose we have a general many-body system, with components labeled
$1,\ldots,N$.  We can regard these labels as a set of vertices, $V$,
for a graph.  Given a two-body Hamiltonian for that system, we can
naturally associate to each coupling between bodies an (undirected)
edge between the corresponding vertices.  So, for example, the
Hamiltonian\footnote{We use $I,X,Y,Z$ to denote the four Pauli
  matrices, and omit tensor product signs for notational brevity.} $H
= XXI+ZIZ$ corresponds to a graph with vertices $1,2,3$, and edges
$\{1,2\}, \{1,3\}$.

%
%
More generally, if some terms in the Hamiltonian couple more than two
bodies, then we can associate with that Hamiltonian a
\emph{hypergraph}. A hypergraph consists of the set, $V$, of vertices,
together with a collection of \emph{hyperedges}, $E$.  Each hyperedge
in $E$ is just a subset of $V$, and represents a coupling term between
the corresponding systems.  So, for example, the Hamiltonian $H =
XXI+ZIZ+YYZ$ corresponds to a hypergraph with vertices $1,2,3$, and
hyperedges $\{1,2\}, \{1,3\}$ and $\{1,2,3\}$.

%
%
We call a hypergraph $G = (V,E)$ a \emph{coupling topology} when it is
associated with a quantum system in this way.  We say that a
Hamiltonian $H$ \emph{respects the coupling topology} $G$ if every
coupling in $H$ corresponds to a hyperedge in $G$.  We don't require
every hyperedge in $G$ to have a corresponding coupling in $H$.  So,
for example, the three-qubit Hamiltonian $H = XXI+ZIZ$ respects the
coupling topology of the hypergraph with vertices $1,2,3$, and
hyperedges $\{1,2\}, \{1,3\}$ and $\{1,2,3\}$, even though there is no
term coupling qubits $1, 2$ and $3$ simultaneously.

%
%
Note that there is an apparent ambiguity in this definition, since a
given Hamiltonian can be decomposed in more than one way, e.g., $H =
XXI + IXX = XXM_+ + M_-XX$, where $M_{\pm} \equiv I\pm X$. We resolve
this ambiguity by saying that $H$ respects the coupling topology $G$
if there is \emph{some} decomposition of $H$ which respects that
coupling topology.

%
%
With this language we can now give a precise statement of the problem
we are interested in.  In fact, it is useful to consider two different
forms of the problem.  The simpler form is as follows:

\textbf{Exact ground-state problem:} Let $\psi$ be a quantum state of
some many-body system.  We think of $\psi$ as a \emph{target state}
that we desire to be the \emph{exact} ground state.  Suppose the
system Hamiltonian $H$ respects the coupling topology $G = (V,E)$.
Given that $\psi$ is an exact ground state of $H$, what does this
imply about the level spacings of $H$?  In particular, do the coupling
topology, $G$, and the correlations present in $\psi$ imply anything
about the level spacings of the system, \emph{independent} of the
specific details of $H$?

%
%
We will show that the answer to this question is ``yes''.  An example
of the sort of answer we'll give is as follows.  Suppose $\psi$ is an
exact ground state of a Hamiltonian, $H$, respecting the coupling
topology $G$.  Then the ground state of $H$ is at least $m$-fold
degenerate, where $m$ is an integer determined solely by (a) the
coupling topology, and (b) the properties of $\psi$.  In particular,
we will see that $m$ is closely related to \emph{long-range}
correlations in $\psi$, where by long range we mean correlations
between systems not directly coupled by $G$.

%
%
It is important that the degeneracy $m$ is determined solely by
properties of $G$ and $\psi$: the particular details of the
Hamiltonian $H$ do not matter, beyond the topology of the
interactions.  Even given the ability to engineer arbitrary designer
Hamiltonians, the fact that $\psi$ is an exact ground state, and $G$
the coupling topology, guarantees an $m$-fold degeneracy in the ground
state.

%
%
More interesting and general than the study of exact ground states is
the study of how the coupling topology and correlations in $\psi$
affect the ability to \emph{approximate} $\psi$ as a ground state.
This question is captured by the following problem.

\textbf{Approximate ground-state problem:} Let $\psi$ be a quantum
state of some many-body system.  Suppose the system Hamiltonian $H$
respects the coupling topology $G = (V,E)$.  Given that the overlap
between $\psi$ and the ground state is $F \equiv \sqrt{\langle
  \psi|P_0|\psi\rangle}$, where $P_0$ projects onto the ground-state
eigenspace, what does this imply about the level spacings of $H$?  In
particular, do the coupling topology, $G$, the overlap $F$, and the
correlations present in $\psi$ imply anything about the level spacings
of the system, \emph{independent} of the specific details of $H$?

%
%
We'll obtain solutions to this problem similar to those obtained for
the exact ground state problem.  For example, suppose $\psi$ has
overlap $F$ with the ground state of a Hamiltonian, $H$, respecting
the coupling topology $G$.  We'll prove an inequality relating the
gap, $\Delta E$, to the overlap, $F$, and a measure, $C$, of long-range
correlation in the system.  This inequality will enable us to prove
that as $F \rightarrow 1$, the presence of long-range correlations in
the system forces the energy gap to vanish.

%
%
In the next two sections we will obtain several solutions to the
approximate ground-state problem, applicable in different contexts.
Interestingly, one of these solutions --- in some sense, the strongest
--- involves quantum error-correcting codes, as discussed in
Sec.~\ref{sec:qecc}.


\section{General theory}
\label{sec:unification}

%
%
Suppose $H$ is a Hamiltonian respecting the coupling topology $G =
(V,E)$, and $\psi$ is a quantum state having overlap $F$ with the
ground state.  Our key result is a general theorem, proved in this
section, connecting the energy levels of $H$ to the properties of a
set we shall define, labelled $R_G(\psi)$.  $R_G(\psi)$ is defined to
consist of all quantum states, both pure and mixed, which agree with
$\psi$ on the hyperedges in $E$.  That is, $R_G(\psi)$ contains all
states $\rho$ such that $\tr_{\overline e}(\rho) = \tr_{\overline
  e}(|\psi\rangle \langle \psi|)$ for all hyperedges $e$ in $E$, where
$\overline e$ indicates that we trace over all systems \emph{except}
those in $e$.

%
%
It is perhaps not obvious why a theorem connecting the energy levels
of $H$ to $R_G(\psi)$ should tell us anything about the relationship
between those energy levels and long-range correlations.  Remarkably,
however, the properties of $R_G(\psi)$ are intimately connected with
the correlations in $\psi$, and this fact will enable us to make the
desired connections.

%
%
Our presentation strategy in this section is to first prove the
general theorem, and then to explore connections between $R_G(\psi)$
and long-range correlations, applying the general theorem to a variety
of examples.

\subsection{Connection between the energy levels and $R_G(\psi)$}

%
%
In this subsection we prove a general theorem connecting the energy
levels of a system having $\psi$ as its approximate ground state to
$R_G(\psi)$.  We begin by specifying some notation and nomenclature.

%
%
Recall that $P_0$ is the projector onto the ground state eigenspace,
and that the overlap between $\psi$ and the ground state is $F \equiv
\sqrt{\langle \psi| P_0 |\psi\rangle}$.  Assuming that $F > 0$, we
define $|E_0\rangle$ to be the (normalized) ground state onto which
$\psi$ projects.  Explicitly, we define $|E_0\rangle \equiv
P_0|\psi\rangle / \sqrt{\langle \psi| P_0|\psi\rangle}$.  It will be
convenient to label the energy levels as $E_0 \leq E_1 \leq \ldots$,
and to let $E_{\max}$ be the largest energy level.  Note that the
energy levels are not assumed to be distinct, so, for example, if the
ground state is doubly degenerate then we will have $E_0 = E_1$.  We
choose $|E_1\rangle, |E_2\rangle,\ldots$ so that
$|E_0\rangle,|E_1\rangle,\ldots$ forms an orthonormal eigenbasis of
energy eigenstates in the obvious way.  We let $E_{\rm tot} = E_{\max}
-E_0$ be the total energy scale for the system.

%
%
With this nomenclature, we are now ready to proceed to the statement
and proof of our main theorem. The key to the proof of the theorem is
a lemma from linear algebra.  The lemma is easy to state, and the
result is rather obvious, yet all the proofs we are aware of make use
of surprisingly sophisticated ideas.  The result appears to be little
known, but is useful in many contexts.  It appeared as Eq.~(133) in a
set of unpublished lecture notes~\cite{Nielsen99d}.

\begin{lemma} {} \label{lemma:trace}
Let $A$ and $B$ be Hermitian matrices.  Then
\begin{eqnarray}
  \lambda(A)^{\downarrow} \cdot \lambda(B)^{\uparrow} \leq \tr(AB)
  \leq \lambda(A)^{\downarrow} \cdot \lambda(B)^{\downarrow},
\end{eqnarray}
where $\lambda(M)$ denotes the vector whose entries are the
eigenvalues of the matrix $M$, $v^{\downarrow}$ (resp. $v^{\uparrow}$)
is the vector whose entries are the entries of $v$ rearranged into
descending (resp. ascending) order, and $\cdot$ is the Euclidean inner
product.
\end{lemma}

\textbf{Proof:} We work in a basis in which $A$ is diagonal, with its
eigenvalues the diagonal entries of the matrix representation in that
basis.  Then
\begin{eqnarray} \label{eq:lemma-inter}
  \tr(AB) = \sum_j A_{jj} B_{jj} = \lambda(A) \! \cdot \mbox{diag}(B),
\end{eqnarray}
where $\mbox{diag}(B)$ is the vector whose entries are the diagonal
elements of $B$ in this basis.  Elementary results from the theory of
majorization imply that $\mbox{diag}(B) \prec \lambda(B)$, where
$\prec$ denotes the majorization relation\footnote{This result appears
  on page~218 of~\cite{Marshall79a}, as Theorem~B.1 in Chapter~9.
  See, e.g., any
  of~\cite{Nielsen99d,Nielsen01b,Bhatia97a,Marshall79a,Alberti82a} for
  an introduction to majorization and further references.}.  Further
elementary results from the theory of majorization\footnote{See
  Page~113 of~\cite{Marshall79a}, Proposition~C.1 of Chapter~4.} imply
that $\mbox{diag}(B) = \sum_j p_j P_j \lambda(B)$, where the $p_j$
form a probability distribution, and the $P_j$ are permutation
matrices.  Substituting into Eq.~(\ref{eq:lemma-inter}) we obtain
\begin{eqnarray}
  \tr(AB) = \sum_j p_j \lambda(A) \cdot P_j \lambda(B).
\end{eqnarray}
The result now follows from the observation\footnote{A proof of this
  observation may be found as Corollary~II.4.4, on page~49
  of~\cite{Bhatia97a}.} that for any two vectors, $x$ and $y$,
$x^{\downarrow} \cdot y^{\uparrow} \leq x \cdot y \leq x^{\downarrow}
\cdot y^{\downarrow}$.

\textbf{QED}

We are now in position to state and prove our main theorem.  Note,
incidentally, that the proof of the main theorem only makes use of the
first inequality in the statement of Lemma~\ref{lemma:trace}, not the
second inequality.  We included both because both are of interest,
appear to be little known, and virtually no extra work is required to
obtain the second.

\begin{theorem} {} \label{thm:unification}
  Let $H$ be a Hamiltonian respecting the coupling topology $G$.
  Suppose $\psi$ is a state with overlap $F$ with the ground state of
  $H$.  Let $\rho \in R_G(\psi)$ have eigenvalues $\rho_1 \geq \rho_2
  \geq \ldots$.  Then
  \begin{eqnarray}
    \sum_{j=1}^{d-1} (E_j-E_0) \rho_{j+1} \leq (1-F^2) E_{\rm tot},
  \end{eqnarray}
  where $d$ is the dimension of state space.
\end{theorem}

%
%
It is sometimes convenient to write the sum in a slightly different
fashion.  Including a $j = 0$ term makes no difference, since
$E_0-E_0$ vanishes, so the sum may be rewritten $\sum_j
(E_j-E_0)\rho_{j+1}$, with the sum over all possible indices, $j$.

\textbf{Proof:} By definition of $F$ as the overlap between $\psi$ and
the ground state, $|E_0\rangle$, we see that up to an unimportant
global phase,
\begin{eqnarray} \label{eq:general-inter}
  |\psi\rangle = F |E_0\rangle + \sqrt{1-F^2} |E_\perp\rangle,
\end{eqnarray}
where $|E_\perp\rangle$ is orthonormal to $|E_0\rangle$.  We now use
this expression to evaluate the average energy for the state
$|\psi\rangle$.  The first term on the right-hand side of
Eq.~(\ref{eq:general-inter}) contributes $F^2 E_0$ to the energy,
while the second term contributes at most $(1-F^2) E_{\max}$, since
the energy of $|E_\perp\rangle$ is no more than $E_{\max}$.  It
follows that $\langle \psi|H|\psi\rangle \leq F^2 E_0 + (1-F^2)
E_{\max}$.  Rewriting this inequality in terms of $E_{\rm tot} =
E_{\max}-E_0$ rather than $E_{\max}$, we obtain
\begin{eqnarray} 
  \langle \psi|H|\psi\rangle \leq E_0 + (1-F^2) E_{\rm tot}.
\end{eqnarray}
Furthermore, since $\psi$ and $\rho$ have the same reduced density
matrices on hyperedges in the coupling topology, we see that $\tr(\rho
H) = \langle \psi|H|\psi\rangle$, and thus:
\begin{eqnarray} \label{eq:rho-bound}
  \tr(\rho H) \leq E_0 + (1-F^2) E_{\rm tot}.
\end{eqnarray}
Applying the first inequality of Lemma~\ref{lemma:trace} to the
left-hand side of Eq.~(\ref{eq:rho-bound}) gives
\begin{eqnarray}
  \sum_{j=1}^d \rho_j E_{j-1} \leq E_0 + (1-F^2) E_{\rm tot}.
\end{eqnarray}
Using the fact that $\sum_{j=1}^d \rho_j =1$, and doing some
elementary algebra and relabeling of indices, we see that this can be
rewritten in the form $\sum_{j=1}^{d-1} (E_j-E_0) \rho_{j+1} \leq
(1-F^2) E_{\rm tot}$, as we set out to prove.

\textbf{QED}

%
%
An interesting observation related to Theorem~\ref{thm:unification} is
that if $G_1$ and $G_2$ are hypergraphs such that the hyperedges of
$G_1$ are a subset of those of $G_2$, then $R_{G_2}(\psi) \subseteq
R_{G_1}(\psi)$.  This is true because if $\rho$ and $\psi$ agree on
hyperedges in $G_2$ then they must certainly agree on hyperedges in
$G_1$.  It follows that Theorem~\ref{thm:unification} implies
\emph{stronger} constraints on the energy levels for systems whose
coupling topology respects $G_1$ than for systems respecting $G_2$.
Thus, for example, Theorem~\ref{thm:unification} gives stronger
constraints on the energy levels for five spins arranged in a line,
with nearest-neighbour interactions, than for the same spins arranged
into a circle, again with nearest-neighbour interactions.

\subsection{Example applications}

%
%
We now explore some applications of Theorem~\ref{thm:unification},
relating the energy spectrum of a system to the presence of long-range
correlations in the ground state of that system.

\subsubsection{Example: Perfect long-range correlations}

Suppose we have a three-component system, with subsystems labeled $1,
2$ and $3$.  Suppose the coupling topology $G$ is such that systems
$1$ and $2$ may interact, systems $2$ and $3$ may interact, but
systems $1$ and $3$ cannot interact directly.  Note that in this
discussion $1$, $2$, and $3$ may be aggregates --- e.g., systems $1$
and $3$ might be spins on either end of a long linear chain, with
system $2$ the collection of all spins inbetween.  Suppose,
furthermore, that $\psi$ is some quantum state exhibiting
\emph{perfect correlation} between systems $1$ and $3$.  By perfect
correlation, we mean that there is a measurement basis in which a
measurement outcome of $j$ on system $1$ implies, with probability
$1$, a measurement outcome $j$ on system $3$, and conversely.

%
%
As an example of such a situation, $\psi$ could be a product
$\psi_{13} \otimes \psi_2$.  In this case $\psi$ exhibits perfect
correlations if measurements are performed in the Schmidt bases for
systems $1$ and $3$, respectively.

%
%
Another example is states $\psi$ such that when system $2$ is traced
out we get a mixed state of the form $\sum_j p_j |j \rangle_1 \langle
j|_1 \otimes |j\rangle_3 \langle j|_3$, where $|j\rangle_1$ and
$|j\rangle_3$ are orthonormal bases for systems $1$ and $3$,
respectively.  It is easy to show that such states must have
three-party Schmidt decompositions of the form studied by
Thapliyal~\cite{Thapliyal99a} and Peres~\cite{Peres95b}, i.e., $\psi =
\sum_j \sqrt{p_j} |j\rangle_1|j\rangle_2 |j\rangle_3$, where
$|j\rangle_1$, $|j\rangle_2$ and $|j\rangle_3$ are orthonormal bases
for the respective systems.  An example of such a state is the GHZ
state $|GHZ \rangle = (|000\rangle+ |111\rangle)/\sqrt 2$.  Indeed, if
we consider an $n$-qubit linear array, with the first and last qubits
considered as system $1$ and system $3$, with the remaining qubits
grouped together as system $2$, then we see that the $n$-party GHZ
state $|GHZ \rangle = (|0\rangle^{\otimes n}+ |1\rangle^{\otimes
  n})/\sqrt 2$ is also an example of such a state.

%
%
In general, if $\psi$ is any state exhibiting such perfect
correlations there must exist normalized, but possibly non-orthogonal,
states $|j\rangle_2$ of system $2$, such that
\begin{eqnarray}
  \psi = \sum_j \sqrt{p_j} |j\rangle_1 |j\rangle_2 |j\rangle_3.
\end{eqnarray}
Note that $p_j$ are the probabilities with which the measurement
outcome $j$ occurs on systems $1$ and $3$.  Now define
\begin{eqnarray}
 \rho \equiv \sum_j p_j |j\rangle_1 \langle j|_1 \otimes |j\rangle_2
 \langle j|_2 \otimes |j\rangle_3 \langle j|_3
\end{eqnarray}
Observe that $\rho \in R_G(\psi)$, since it has the same reduced
density matrices on systems $12$ and $23$ as does $\psi$.  Note also
that $\rho$ has eigenvalues $p_j$.  It will be convenient to assume
that the measurement outcomes are labelled $1,2,\ldots,d$, and have
been ordered so that $p_1 \geq p_2 \geq \ldots$.  From
Theorem~\ref{thm:unification} we have
\begin{eqnarray} \label{eq:perfect-correlation}
  \sum_{j=1}^{d-1} (E_j-E_0) p_{j+1}
  \leq (1-F^2) E_{\rm tot}.
\end{eqnarray}

%
%
Eq.~(\ref{eq:perfect-correlation}) tells us that as $F \rightarrow 1$,
the quantity on the left-hand side gets squeezed toward zero.  In
particular, if $p_1,\ldots, p_k > 0$, then we conclude that $E_{k-1}
\rightarrow E_0$, as do all the lower energy levels,
$E_1,\ldots,E_{k-2}$.  So, for example, in the scenario of
Sec.~\ref{sec:toy-model}, if $\psi = |{\rm
  ME}\rangle_{13}|\phi\rangle_2$, then we have $p_1 = p_2 = \frac 12$,
and Eq.~(\ref{eq:perfect-correlation}) becomes\footnote{C.f.
  Eq.~(\ref{eq:intro-example}).} $E_1 \leq E_0 + 2(1-F^2) E_{\rm
  tot}$.  Thus, the gap to the first excited state in this example
vanishes as $F \rightarrow 1$.

%
%
Similarly, for an $n$-qubit linear chain, with systems $1$ and $3$ the
qubits on each end of the chain, if $\psi$ is the $n$-party GHZ state,
then we again conclude that $E_1 \leq E_0 + 2(1-F^2) E_{\rm tot}$, and
the gap to the first excited state vanishes as $F \rightarrow 1$.

%
%
Another illuminating --- albeit, ultimately trivial --- example is
when system $13$ is in a product state, $\psi =
|a\rangle_1|b\rangle_2|c\rangle_3$.  This is a case of the theorem,
for the system does exhibit perfect correlation, provided system $1$
is measured in a basis including $|a\rangle$, and system $3$ is
measured in a basis including $|b\rangle$.  However, since we have
$p_1 = 1$, and all other $p_j = 0$, we see that
Eq.~(\ref{eq:perfect-correlation}) gives us only trivial information,
$0 \leq (1-F^2) E_{\rm tot}$, and cannot be used to deduce anything
about the spectrum of the system.  It is only as the probabilities
$p_j$ become mixed that Eq.~(\ref{eq:perfect-correlation}) may be used
to deduce interesting information about the spectrum.

\subsubsection{Example: Imperfect long-range correlations}

Let us generalize the previous example so that it applies also to
systems with imperfect correlations.  Suppose again that we have a
three-component system, $123$, and the coupling topology allows $1$
and $2$, and $2$ and $3$ to interact, but not $1$ and $3$.  Suppose
$\psi$ is an \emph{exact} ground state for a Hamiltonian respecting
this coupling topology.  Suppose $|j\rangle$ is an orthonormal basis
for system $1$, and $|k\rangle$ is an orthonormal basis for system
$3$.  We can expand $\psi$ as:
\begin{eqnarray}
  \psi = \sum_{jk} \sqrt{p_{j,k}} |j\rangle |e_{jk}\rangle |k\rangle,
\end{eqnarray}
where $p_{j,k}$ is the probability of getting the measurement outcome
$j$ on system $1$ and $k$ on system $3$, if measurements are performed
in the $|j\rangle$ and $|k\rangle$ bases, respectively.  The states
$|e_{jk}\rangle$ are normalized, but possibly non-orthogonal, states
of system $2$.

%
%
To measure the correlation between the measurement outcomes on systems
$1$ and $3$ we define a correlation measure,
\begin{eqnarray}
  C \equiv \sum_j p_{j,j}.
\end{eqnarray}
$C$ is just the probability that the measurement outcome on system $1$
is the same as the measurement outcome on system $3$.  Thus, values of
$C$ close to one indicate highly correlated measurement outcomes,
while values very close to zero indicate a high level of
anti-correlation.  

%
%
The definition of $C$ implicitly assumes that the same labels, $j$,
are being used for measurement outcomes on system $1$ and system $3$.
This need not be the case.  For example, system $1$ might be a
spin-$\frac{1}{2}$ system, with measurement outcomes labeled $\pm
\frac 12$, and system $3$ a spin-$1$ system, with measurement outcomes
labeled $0,\pm 1$.  If this is the case we can define an analogous
notion of correlation by identifying the outcomes of the spin-$\frac
12$ measurement with a subset of the spin-$1$ outcomes, e.g., $1/2
\rightarrow 1, -1/2 \rightarrow -1$, and so $C = p_{1/2, 1} +p_{-1/2,
  -1}$.  In general, we can define a measure of correlation by
identifying the measurement outcomes for the system with the smaller
state space with a subset of the measurement outcomes for the system
with the larger state space.  The arguments below are easily
generalized to this case, but for notational clarity we stick to the
case when systems $1$ and $3$ have identical labelings for their
measurements.

%
%
Next, we define a normalized and perfectly correlated state, $\psi'$,
of the joint system by discarding those terms in $\psi$ that lead to
the correlations being imperfect, and renormalizing the state
appropriately:
\begin{eqnarray}
  \psi' \equiv \frac{\sum_j \sqrt{p_{j,j}} |j\rangle |e_{jj}\rangle
    |j\rangle}{\sqrt{C}}.
\end{eqnarray}
$\psi'$ obviously exhibits perfect correlation between systems $1$ and
$3$, in the sense of the earlier example, and thus we conclude that
\begin{eqnarray}
  \sum_{j}  \frac{p_{j+1,j+1}}{C} (E_j-E_0) \leq (1-F^2) E_{\rm tot},
\end{eqnarray}
where $F$ is the overlap between $\psi'$ and the ground state.  But we
assumed that $\psi$ was a ground state (possibly one of many), so $F
\geq |\langle \psi'|\psi\rangle| = \sqrt{C}$, and thus the previous
equation may be rewritten
\begin{eqnarray} \label{eq:correlation-bound}
  \sum_{j} p_{j+1,j+1} (E_j-E_0) \leq C(1-C) E_{\rm tot},
\end{eqnarray}

%
%
Eq.~(\ref{eq:correlation-bound}) tells us that as $C \rightarrow 1$,
i.e., as we approach perfect correlation, the quantity on the
left-hand side must approach zero.  Thus, if $p_{1,1},\ldots,p_{k,k} >
0$ then we conclude that $E_1,\ldots,E_{k-1} \rightarrow E_0$ as the
correlations become perfect.  Interestingly,
Eq.~(\ref{eq:correlation-bound}) also tells us that the same
phenomenon occurs as $\psi$ becomes perfectly anti-correlated, i.e.,
as $C \rightarrow 0$.  Physically, this means that a measurement
outcome of $j$ on system $1$ implies that a measurement outcome
different from $j$ occurred on system $3$.

\subsubsection{Example: Approximating a state with imperfect long-range
correlations}

%
%
We can generalize the previous two examples still further, to the case
where we are trying to \emph{approximate} a state with imperfect
correlations as the ground state.  Suppose again that we have a
three-component system, $123$, and the coupling topology allows $1$
and $2$, and $2$ and $3$ to interact, but not $1$ and $3$.  Suppose
$\psi$ is a state with correlation $C = \sum_j p_{j,j}$ in some
measurement basis for systems $1$ and $3$.  Suppose there is a
Hamiltonian respecting the coupling topology such that the overlap
between $\psi$ and the ground state is $F$.  We will prove that the
energy levels of the Hamiltonian satisfy
\begin{eqnarray}
  & & \sum_{j}
  p_{j+1,j+1} (E_j-E_0) \nonumber \\
  & \leq & C \left( \sqrt{1-C}+\sqrt{1-F^2}\right)^2
  E_{\rm tot}.
\end{eqnarray}
This result generalizes both the last example,
Eq.~(\ref{eq:correlation-bound}), which corresponds to the case when
$F = 1$, and the example before that,
Eq.~(\ref{eq:perfect-correlation}), which corresponds to the case $C =
1$.

%
%
Similarly to the previous example, we can write $\psi = \sum_{jk}
\sqrt{p_{j,k}} |j\rangle |e_{jk}\rangle |k\rangle$, and define $\psi'
\equiv \sum_{j} \sqrt{p_{j,j}} |j\rangle |e_{jj}\rangle |j\rangle /
\sqrt{C}$.  We now define $F(a,b) \equiv |\langle a|b\rangle|$, the
overlap between any two states $|a\rangle$ and $|b\rangle$.  It is
convenient to note that $\sqrt{1-F(a,b)^2}$ is a metric on projective
state space.  Recall that $|E_0\rangle = P_0|\psi\rangle /
\sqrt{\langle \psi| P_0|\psi\rangle}$ is the normalized state that
arises from projecting $\psi$ onto the ground state.  From the
triangle inequality
\begin{eqnarray} 
  \sqrt{1-F(\psi',E_0)^2} & \leq & \sqrt{1-F(\psi',\psi)^2}
  + \sqrt{1-F(\psi,E_0)^2} \nonumber \\
  & & \\
  & \leq & \sqrt{1-C} + \sqrt{1-F^2}. \label{eq:approx-imperfect-correlation}
\end{eqnarray}
But if $F_{\psi'} \equiv \sqrt{\langle \psi'|P_0|\psi'\rangle}$ is the
overlap of $\psi'$ with the ground state then we have $F_{\psi'} \geq
F(\psi',E_0)$, and thus, combining with
Eq.~(\ref{eq:approx-imperfect-correlation}) we have
\begin{eqnarray} 
  1-F_{\psi'}^2 \leq \left( \sqrt{1-C} + \sqrt{1-F^2} \right)^2.
\end{eqnarray}
The result now follows from Eq.~(\ref{eq:perfect-correlation}).

%
%
Summarizing, we have proved the following general theorem:
\begin{theorem} {} \label{thm:gen-corr-bound}
  Let $H$ be a Hamiltonian coupling systems $1$ and $2$, and $2$ and
  $3$, but not $1$ and $3$.  Let $p_{j,k}$ be the joint probability
  distribution associated to a measurement in some bases for systems
  $1$ and $3$, for a state $\psi$.  Label the measurement outcomes
  $1,2,\ldots$, and so that $p_{1,1} \geq p_{2,2} \geq \ldots$.
  Define the correlation measure $C \equiv \sum_j p_{j,j}$, and
  let $F$ be the overlap between $\psi$ and the ground state.  Then
  the energy levels of $H$ are constrained by the relation:
\begin{eqnarray}
  & & \sum_{j}
  p_{j+1,j+1} (E_j-E_0) \nonumber \\
  & \leq & C \left( \sqrt{1-C}+\sqrt{1-F^2}\right)^2
  E_{\rm tot}.
\end{eqnarray}
\end{theorem}

\subsection{Exact ground states and ground-state degeneracy}
\label{subsec:exact-case}

%
%
We've seen that the properties of $R_G(\psi)$ are closely related to
long-range correlations in the state $\psi$.  In this section we make
some more specialized observations about $R_G(\psi)$ that can be used
to prove results about the ground-state degeneracy of any Hamiltonian
with $\psi$ as an \emph{exact} ground state.

%
%
We define $N_{\rm rank}(\psi)$ to be the maximal rank of any density
matrix in $R_G(\psi)$.  We will see below that $N_{\rm rank}(\psi)$ is
connected to both the long-range correlations in $\psi$, and also to the 
ground-state degeneracy.  We begin with the latter connection:
\begin{theorem} {} \label{thm:exact}
  Let $H$ be a Hamiltonian respecting the coupling topology $G$.
  Suppose $\psi$ is a ground state of $H$.  Then the ground state is
  at least $N_{\rm rank}(\psi)$-fold degenerate.
\end{theorem}

%
%
\textbf{Proof:} A direct proof is easily obtained.  Let $\rho$ be the
state in $R_G(\psi)$ of maximal rank, let $\psi_j$ be the eigenvectors
of $\rho$ with non-zero eigenvalues, and argue that all the $\psi_j$
must have energy equal to the ground state energy.  This follows since
if one has energy higher than the ground state, then another must have
energy below the ground state --- a contradiction --- to ensure that
$\mbox{tr}(H\rho)$ is equal to the ground state energy.  Alternately,
observe that this theorem is a special case of
Theorem~\ref{thm:unification}, with $F=1$.

\textbf{QED}

%
%
\textbf{Example:} As an example, suppose we have just three systems,
$1, 2, 3$, and suppose only couplings between $12$ and $23$ are
involved.  Suppose that $\psi = \psi_{13} \otimes \psi_2$, where
$\psi_{13}$ is an entangled state of systems $1$ and $3$, with Schmidt
decomposition $\psi_{13} = \sum_j \sqrt{p_j}|j\rangle|j\rangle$, and
$\psi_2$ is some state of system $2$.

%
%
We will analyse this scenario in two different ways.  The first method
of analysis is similar in spirit to arguments earlier in the paper,
such as led to Theorem~\ref{thm:gen-corr-bound}.  The second method is
from a somewhat different point of view, and we will see that it
sometimes leads to stronger results.  Our first argument is as
follows.  Just as argued earlier, $\rho = \sum_j p_j |j\rangle \langle
j| \otimes |\psi_2\rangle \langle \psi_2| \otimes |j\rangle \langle
j|$ is in $R_G(\psi)$.  We therefore see, from any one of
Theorems~\ref{thm:exact},~\ref{thm:gen-corr-bound},~\ref{thm:unification},
that the ground-state degeneracy is at least equal to the Schmidt
number of $\psi_{13}$, $\mbox{Sch}(\psi_{13})$, i.e., the number of
non-zero coefficients in the Schmidt decomposition.  It follows that
if $\psi_{13} \otimes \psi_2$ is to be a ground state of the system,
then the ground state must be $\mbox{Sch}(\psi_{13})$-fold degenerate.
Of course, the Schmidt number is a well-known entanglement monotone,
so in this example we conclude that the ground-state degeneracy is at
least as large as the amount of long-range entanglement, as measured
by the Schmidt number.

%
%
Our second method of analysis takes a state-based, rather than
operator-based, point of view.  Let $S_G(\psi)$ be the set of
\emph{pure} quantum states agreeing with $\psi$ on hyperedges, i.e.,
it is the subset of $R_G(\psi)$ containing only pure states.  Define
$N_{\rm span}(\psi)$ to be the dimension of the linear space spanned
by the vectors in $S_G(\psi)$.  Observe then that $N_{\rm span}(\psi)
\leq N_{\rm rank}(\psi)$, since given any linearly independent
$\psi_1,\ldots,\psi_m \in S_G(\psi)$ we can form $\rho = \sum_j
|\psi_j\rangle \langle \psi_j|/m \in R_G(\psi)$, which has rank $m$.
Thus, Theorem~\ref{thm:exact} implies that the ground state is at
least $N_{\rm span}(\psi)$-fold degenerate.

%
%
In the scenario studied above, with $\psi = \psi_{13} \otimes \psi_2$,
$\psi_{13} = \sum_j \sqrt{p_j} |j\rangle |j\rangle$, we see that the
states $\sum_j \sqrt{p_j}e^{i \theta_j}|j\rangle |j\rangle$ are in
$S_G(\psi)$ for any choice of the phases $\theta_j$, and thus $N_{\rm
  span}(\psi) \geq \mbox{Sch}(\psi_{13})$, and we conclude, as
earlier, that the ground state is at least
$\mbox{Sch}(\psi_{13})$-fold degenerate.  However, when the Schmidt
coefficients $p_j$ are degenerate, $N_{\rm span}(\psi)$ can actually
be somewhat larger than the Schmidt number $\mbox{Sch}(\psi_{13})$.
The following proposition enables us to make a precise evaluation of
$N_{\rm span}(\psi)$.

\begin{proposition}
  Let $\psi = \psi_{13} \otimes \psi_2$, where $\psi_{13} = \sum_j
  \sqrt{p_j} |j\rangle |j\rangle$.  Then $N_{\rm span}(\psi) = \sum_k
  d_k^2$, where the sum is over an index $k$ for \emph{distinct}
  non-zero Schmidt coefficients, and $d_k$ is the degeneracy of the
  $k$th non-zero Schmidt coefficient.
\end{proposition}

Note that, according to the proposition, when $\psi_{13}$ has
non-degenerate Schmidt coefficients, $N_{\rm span}(\psi)$ is equal to
the Schmidt number of $\psi_{13}$, which is an entanglement
monotone.  However, using the results of~\cite{Nielsen99a} it is easy
to construct examples with degenerate Schmidt coefficients that show
$N_{\rm span}(\psi)$ is not, in general, an entanglement monotone.

\textbf{Proof:} It is clear that all states in $S_G(\psi)$ have the
form $\phi_{13} \otimes \psi_2$ where $\phi_{13}$ is a state having
the same reduced density matrices on systems $1$ and $3$ as does
$\psi_{13}$.  But it is easy to see that this is the case if and only
if $\phi_{13} = e^{i \theta} \left( (\oplus_k U_k) \otimes I\right)
\psi_{13}$, where $\theta$ is a phase factor, $U_k$ is a special
unitary operator acting on the subspace of system $1$ corresponding to
the $k$th Schmidt coefficient, and $\oplus_k$ denotes the direct sum
over those subspaces.  The result now follows from the simple
observation that in a $d_k \otimes d_k$ space, the dimension spanned by
states of the form $(U \otimes I) \sum_j |j\rangle |j\rangle$, where
$U \in SU(d_k)$, is $d_k^2$.

\textbf{QED}

%
%
This proposition shows us how to evaluate $N_{\rm span}(\psi)$ for a
large class of interesting states, and thus to place lower bounds on
the ground-state degeneracy.  When $\psi_{13}$ is degenerate these
results are actually stronger than are obtained using
Theorem~\ref{thm:gen-corr-bound}, since $N_{\rm span}(\psi)$ is
strictly larger in this case than the Schmidt number of $\psi_{13}$.
Although the argument leading to Theorem~\ref{thm:gen-corr-bound} can
be modified to give this stronger bound, the modification is not
especially natural from a physical point of view.  Thus, we believe
there is some merit in the alternate, state-based point of view taken
in the present discussion.

\textbf{Example:} Recall that a state with a multi-party Schmidt
decomposition can be written in the form~\cite{Peres95b,Thapliyal99a}
$\psi = \sum_j \sqrt{p_j} |j\rangle|j\rangle \ldots |j\rangle$.  An
example of such a state is the $n$-qubit GHZ state $|GHZ \rangle =
(|0\rangle^{\otimes n} + |1\rangle^{\otimes n})/\sqrt 2$.  Suppose the
coupling topology $G$ contains all hyperedges of up to $n-1$ vertices,
i.e., the allowed Hamiltonians may couple up to $n-1$ of the systems,
but not all $n$ systems simultaneously.  It is easy to see that the
states $\sum_j \sqrt{p_j}e^{i \theta_j}|j\rangle \ldots |j\rangle$ are
in $S_G(\psi)$, for any choice of the phases $\theta_j$, and thus
$N_{\rm span}(\psi) \geq \mbox{Sch}(\psi)$, where $\mbox{Sch}(\psi)$
is the number of terms appearing in the multi-party Schmidt
decomposition.  It follows that the ground state of $H$ is at least
$\mbox{Sch}(\psi)$-fold degenerate.  For example, in the case of the
GHZ state, it follows that the ground state is at least two-fold
degenerate, since the GHZ state has Schmidt number two.

\subsection{Further development of Theorem~\ref{thm:unification}}

%
%
Can Theorem~\ref{thm:unification} be strengthened in any way?  We now
show that there are physically interesting ways of varying the
hypotheses of Theorem~\ref{thm:unification}, in order to reach
stronger conclusions.  One way way of doing this, related to quantum
error-correcting codes, is described in detail in
Section~\ref{sec:qecc}.  We now explain, more briefly, another
possible variation.

%
%
The basic idea is to amend Theorem~\ref{thm:unification} so it makes
use of information about the relationship between $\psi$ and $\rho$.
Consider two possible cases: (a) $\psi$ is orthogonal to the support
of $\rho$, and (b) $\psi$ is contained in the support of $\rho$.  In
the former case, we see that there is a subspace of dimension
$\mbox{rank}(\rho)+1$, spanned by the support of $\rho$ and $\psi$, in
which energies are all approximately equal to $E_0$, and thus $E_0
\approx E_1 \approx \ldots E_{\mbox{rank}(\rho)}$.  In the latter case
we can only conclude that there is a subspace of dimension
$\mbox{rank}(\rho)$ --- the support of $\rho$ --- in which energies
are all approximately equal to $E_0$, and thus we draw the weaker
conclusion that $E_0 \approx E_1 \approx \ldots
E_{\mbox{rank}(\rho)-1}$.

%
%
We have not yet succeeded in obtaining a clean generalization of
Theorem~\ref{thm:unification} incorporating this idea.  However, we
have obtained a simpler result in this vein, which we now briefly
describe.

\begin{proposition} {} \label{prop:trial_propn}
  Let $H$ be a Hamiltonian respecting the coupling topology $G$.
  Suppose $\psi$ is a state having overlap $F$ with the ground state
  of $H$.  Suppose $\phi \in R_G(\psi)$ is such that $|\langle
  \psi|\phi \rangle| = \cos(\theta)$.  Then
  \begin{eqnarray} \label{eq:special-bound}
    E_1 -E_0 \leq \frac{1-F^2}{g(\theta,F)}E_{\rm tot},
  \end{eqnarray}
  where $g(\theta,F) \equiv
  1-(F\cos(\theta)+\sqrt{1-F^2}\sin(\theta))^2$.
\end{proposition}

Note that $\phi$ plays a role analogous to $\rho$ in
Theorem~\ref{thm:unification}.  The crucial additional piece of
structure in the proposition is the angle $\theta$ relating $\psi$ and
$\phi$.  As this angle varies from $0$ to $\pi/2$, the bound
Eq.~(\ref{eq:special-bound}) varies from the vacuous $E_1 - E_0 \leq
E_{\rm tot}$ --- as with Theorem~\ref{thm:unification} we get no
information at all in this case --- through to $E_1 - E_0 \leq
(1-F^2)E_{\rm tot}/F^2 $, which is non-trivial.  Note that
Theorem~\ref{thm:unification} can be applied also in this latter case;
the strongest bound obtained in this way comes from choosing $\rho =
\frac{1}{2} |\psi\rangle \langle \psi|+\frac{1}{2} |\phi\rangle
\langle \phi|$, which gives $E_1 -E_0 \leq 2 (1-F^2) E_{\rm tot}$,
which is a factor of two weaker than
Proposition~\ref{prop:trial_propn}, in the $F \rightarrow 1$ limit.

\textbf{Proof:} By the same argument that led to
Eq.~(\ref{eq:rho-bound}), we conclude that
\begin{eqnarray} \label{eq:phi-bound}
  \langle \phi|H|\phi\rangle \leq E_0+(1-F^2) E_{\rm tot}.
\end{eqnarray}
Expressing $|E_0\rangle$ in terms of $\psi$ we have, up to an
unimportant global phase, $|E_0\rangle = F|\psi\rangle+\sqrt{1-F^2}
|\psi^\perp\rangle$, for some $\psi^\perp$ orthonormal to $\psi$.
Taking the inner product with $\phi$ gives $|\langle \phi|E_0\rangle |
\leq F \cos \theta +\sqrt{1-F^2} |\langle \phi|\psi^\perp\rangle|$.
Because $\psi^\perp$ is orthonormal to $\psi$ we have $|\langle
\phi|\psi^\perp\rangle| \leq \sin \theta$, and so
\begin{eqnarray}
  |\langle \phi|E_0\rangle | \leq F \cos \theta +\sqrt{1-F^2} 
  \sin \theta.
\end{eqnarray}
We see from this equation that the component of $\phi$ orthogonal to
$|E_0\rangle$ is at least $\sqrt{g(\theta,F)}$, as defined in the
statement of the proposition, and thus
\begin{eqnarray}
  \langle \phi|H|\phi\rangle \geq (1-g(\theta,F)) E_0 + g(\theta,F)
E_1.
\end{eqnarray}
Combining this inequality with Eq.~(\ref{eq:phi-bound}) and
rearranging gives the result.

\textbf{QED}

\subsection{Understanding $R_G(\psi)$}

%
%
The key to applying Theorem~\ref{thm:unification} is the ability to
find states $\rho$ lying in $R_G(\psi)$.  To this end, we make a few
general remarks on the problem of understanding $R_G(\psi)$.

%
%
Our first observation is that $R_G(\psi)$ is a convex set, since a
mixture of states, each of which agrees with $\psi$ on hyperedges,
also agrees with $\psi$ on hyperedges.  Therefore, one might try to
understand $R_G(\psi)$ by finding its extreme points.  Unfortunately,
we do not know what those extreme points are, or even if they are pure
or mixed quantum states.

%

%
%
Additional light on $R_G(\psi)$ is shed by the work of Linden, Popescu
and Wootters~\cite{Linden02a}, and subsequent work by Linden and
Wootters~\cite{Linden02b}.  In~\cite{Linden02a} it is shown that
almost all three-qubit quantum states are uniquely determined by their
two-party reduced density matrices.  More precisely, given a
three-qubit state $\psi = \psi_{123}$, let $\rho_{12}, \rho_{13},
\rho_{23}$ be the corresponding two-qubit reduced density matrices.
Then~\cite{Linden02a} show that unless the state is equivalent, up to
local unitaries, to a state of the form $a|000\rangle +b|111\rangle$,
then $\psi$ is the unique state, even allowing mixed states, with
those reduced density matrices.

%
%
Restating in our language,~\cite{Linden02a} shows that for all $\psi$
except those equivalent to $a|000\rangle+b|111\rangle$ by local
unitaries, $R_G(\psi) = \{ \psi \}$, when $G$ is the complete graph
allowing interactions between any pair of the systems $1, 2$ and $3$.
Thus, Theorem~\ref{thm:unification} only gives non-trivial information
when the state $\psi$ is locally equivalent to
$a|000\rangle+b|111\rangle$.  Of course, bounds like
Theorem~\ref{thm:gen-corr-bound} apply in general.

%
%
The results of~\cite{Linden02a} were extended in~\cite{Linden02b},
which considered the scenario of $n$ qudits, i.e., $d$-dimensional
quantum systems.  \cite{Linden02b} proved the existence of constants
$\alpha$ and $\beta$, $0 < \alpha < \beta < 1$ such that: (a)
specifying all reduced density matrices for subsystems containing
$\beta n$ qudits uniquely determined the global state for almost all
quantum states, and (b) knowing all the reduced density matrices on up
to $n \alpha$ qudits does \emph{not} uniquely determine the state, in
general.  The estimates they obtained for $\alpha$ and $\beta$ were of
order $1$, and depended on the value of $d$; for details,
see~\cite{Linden02b}.

%
%
Restating in our language,~\cite{Linden02b} showed that if $G$
includes all hyperedges involving up to $\beta n$ vertices, then for
almost all $\psi$, $R_G(\psi) = \{ \psi \}$.  However, for more
physically interesting cases, like when the coupling topology only
involves two-body interactions, the results of~\cite{Linden02b}
suggest that $R_G(\psi)$ will typically contain mixed states, and thus
the bounds of Theorem~\ref{thm:unification} become non-trivial.

\section{Connection to quantum error-correcting codes}
\label{sec:qecc}

%
%
There is an interesting way to strengthen the conclusions of the
earlier theorems, by making use of stronger hypotheses.  Intriguingly,
this line of thinking leads to a natural connection with quantum
error-correcting codes.  We present this material starting with a
general theorem connecting the gap to the properties of the ground
state, and then explain how those properties are connected to quantum
error-correcting codes.

%
%
We begin with a little more notation.  Let $\overline S_G(\psi)$
denote the set of all vectors $\lambda \phi$, where $\lambda$ is a
complex number, and $\phi$ is a state in $S_G(\psi)$.  Let $N_{\rm
  space}(\psi)$ be the dimension of the largest vector space which is
a subset of $\overline S_G(\psi)$.  We now prove that $N_{\rm
  space}(\psi)$ is connected to the spectral properties of the system.

\begin{theorem} {} \label{thm:qecc}
  Let $H$ be a Hamiltonian respecting the coupling topology $G$.
  Suppose $\psi$ is a state with overlap $F$ with the ground state.
  Then 
  \begin{eqnarray}
    E_0 \leq E_1 \leq \ldots \leq E_{N_{\rm space}(\psi)-1} \leq
  E_0+ (1-F^2)E_{\rm tot}.
  \end{eqnarray}
\end{theorem}

The inequality which is the conclusion of this theorem is
substantially stronger than the inequalities proved earlier, such as
Theorem~\ref{thm:unification} and its corrolaries.  The reason this
stronger conclusion is possible is because we use a stronger
hypothesis as the basis for our reasoning.  The key fact is that every
state in the maximal subspace of $\overline S_G(\psi)$ is guaranteed
to have the same expectation energy for Hamiltonians respecting $G$.
In contrast, in the scenario of Theorem~\ref{thm:unification}, we know
that $\rho \in R_G(\psi)$, but this does not imply that all states in
the support of $R_G(\psi)$ have the same expectation energy.  It is
this difference that allows us to draw a stronger conclusion in the
present scenario.

\textbf{Proof:} Let $V$ be the maximal vector space which is a subset
of $\overline S_G(\psi)$.  By the Courant-Fischer-Weyl minimax
principle (see Chapter~3 of~\cite{Bhatia97a}), we have
\begin{eqnarray}
E_{N_{\rm space}(\psi)-1} 
& \leq & \max_{\phi \in V, \|\phi\| = 1} \langle \phi|H|\phi\rangle.
\end{eqnarray}
But by the same reasoning that led to Eq.~(\ref{eq:rho-bound}) the
right-hand side of the previous equation is bounded above by $E_0 +
(1-F^2) E_{\rm tot}$, which gives the result.

\textbf{QED}

%
%
How can we evaluate $N_{\rm space}(\psi)$?  Insight into this question
is provided by noticing an interesting connection, namely, that the
maximal vector space contained in $\overline S_G(\psi)$ is a type of
quantum error-correcting code.  To see this, let us recall some basic
facts from the theory of quantum
error-correction~\cite{Nielsen00a,Preskill98c}.

%
%
Let $S$ be a set whose elements are collections of subsystems of some
quantum system.  The elements of $S$ represent (collective) subsystems
on which errors are allowed to occur, and still be correctable by the
code.  For example, for a code correcting errors on up to two qubits
at a time, $S$ consists of all pairs $\{j,k\}$ of labels for two
qubits.  A quantum error-correcting code correcting errors on $S$ is a
vector space, $W$, such that
\begin{eqnarray} \label{eq:qecc}
  P A^\dagger B P \propto P,
\end{eqnarray}
where $P$ projects onto the code space $W$, and $A$ and $B$ are
arbitrary operators that act non-trivially only on subsystems which
are elements of $S$.  These conditions, Eq.~(\ref{eq:qecc}), define
what it is to be a quantum error-correcting code correcting errors on
$S$.  For more on the physical interpretation of these conditions,
see~\cite{Nielsen00a,Preskill98c}.

%
%
We return now to the connection between Theorem~\ref{thm:qecc} and
quantum error-correction.  In one direction, the connection is quite
simple.  Suppose $\psi$ is a state in a $k$-dimensional quantum
error-correcting code, $W$, which corrects errors on a set, $S$.  We
define a coupling topology on the system, $G = (V,E)$, by specifying
that $E$ consists of all hyperedges $e$ such that $e\subseteq s_1 \cup
s_2$ for some $s_1, s_2 \in S$.  We will use Eq.~(\ref{eq:qecc}) to
show that all states $\phi$ in the code, $W$, must have the same
reduced density matrices on any hyperedge $e$, and thus $W \subseteq
\overline S_G(\psi)$, and therefore $N_{\rm ortho}(\psi) \geq k$.

%
%
To see this, suppose $C$ is an operator which is a tensor product of
operators acting on the individual systems in $e$.  It follows that
$C= A^\dagger B$ for some operators $A$ and $B$ acting only on the
systems in $s_1$ and $s_2$.  We have, by Eq.~(\ref{eq:qecc}), $P C P =
\gamma P$ for some constant of proportionality $\gamma$.  It follows
that if $\phi$ is any state in the code then
\begin{eqnarray}
 \tr(|\phi\rangle \langle \phi| C) = \gamma
\end{eqnarray}
This is true for all $\phi$ in the code, and because $C$ was an
arbitrary tensor product acting on $e$, we see that the reduced
density matrix on $e$ must be the same for all elements $\phi$ of the
code.

%
%
The converse statement is also true.  Suppose $W$ is the maximal
subspace in $\overline S_G(\psi)$.  Suppose $S$ is any set such that
for each pair $s_1$ and $s_2$ in $S$ there is a hyperedge $e$ in $E$
satisfying $e \supseteq s_1 \cup s_2$.  We will show that $W$ is an
error-correcting code correcting errors on $S$.  The proof is similar
to but slightly more elaborate than the proof in the previous
paragraph.  Let $A$ and $B$ be operators acting non-trivially only on
subsystems $s_1$ and $s_2$.  We aim to establish Eq.~(\ref{eq:qecc}).
Because all states $\phi$ in $S_G(\psi)$ have the same reduced density
matrices on $e$ we conclude that
\begin{eqnarray}
  \langle \phi| A^\dagger B |\phi\rangle = \gamma,
\end{eqnarray}
for some constant $\gamma$ independent of $\phi$.  This implies 
\begin{eqnarray} \label{eq:qecc-inter}
  |\phi\rangle \langle \phi| A^\dagger B |\phi\rangle \langle \phi|= \gamma
  |\phi\rangle \langle \phi|.
\end{eqnarray}
Naively, one might try to establish Eq.~(\ref{eq:qecc}) by summing
over an orthonormal basis of states $\phi$ for $W$.  Of course, this
may not work, because of possible cross terms on the left-hand side of
Eq.~(\ref{eq:qecc-inter}).  We will show, however, that these
cross-terms vanish.  To see this, let $|j\rangle$ be an orthonormal
basis for $W$.  Then for any pair $j_1 \neq j_2$ we have
\begin{eqnarray}
& & \left(\langle j_1|+\langle j_2|\right) A^\dagger
  B \left( |j_1\rangle + |j_2\rangle
    \right) \nonumber {} \\
 & = &
  \left(\langle j_1|-\langle j_2|\right) A^\dagger
B \left( |j_1\rangle - |j_2\rangle
    \right)
\end{eqnarray}
and
\begin{eqnarray}
& & \left(\langle j_1|-i\langle j_2|\right) A^\dagger
  B \left( |j_1\rangle+i|j_2\rangle
    \right) \nonumber \\
 & = &
  \left(\langle j_1|+i\langle j_2|\right) A^\dagger
  B \left( |j_1\rangle -i|j_2\rangle
    \right). 
\end{eqnarray}
Adding the first of these equations to $i$ times the second equation gives
\begin{eqnarray}
  \langle j_2|A^\dagger B |j_1\rangle = 0,
\end{eqnarray}
which establishes that the cross-terms vanish, and thus that $W$ is
a quantum error-correcting code.

%
%
We have shown that systems with quantum error-correcting codes as
approximate ground states must satisfy especially stringent
constraints on their low-lying spectra.  It is interesting to compare
these results with those of~\cite{Haselgrove03b}, where it was shown
that non-degenerate quantum error-correcting codes correcting errors
on up to $L$ subsystems cannot be the ground state of any non-trivial
$L$-local Hamiltonian, i.e., a Hamiltonian coupling no more than $L$
subsystems at a time, and not a multiple of the identity.
Remarkably,~\cite{Haselgrove03b} proved a \emph{constant} lower bound
on the distance between the ground state and states of the code in
this scenario.  This constant lower bound is much stronger even than
the bounds of Theorem~\ref{thm:qecc}.  However, a critical difference
is that the results of~\cite{Haselgrove03b} applied only to
non-degenerate codes, while Theorem~\ref{thm:qecc} is more general in
that it applies also to degenerate codes.

Viewed from a slightly different angle, our results provide an amusing
counterpoint to~\cite{Haselgrove03b}.  \cite{Haselgrove03b} pointed
out that no state in a non-degenerate code correcting up to $L$ errors
can be a ground state of an $L$-local Hamiltonian.
Theorem~\ref{thm:qecc} implies that if one state of a degenerate code
correcting $L$ errors is a ground state of an $L$-local Hamiltonian,
then \emph{all states} of that code must be ground states of the
Hamiltonian.  Physically, this is clear \emph{a priori} --- all the
states of the code must be energetically indistinguishable, in order
to preserve information.  However, it seems to us an interesting fact
that either all or none of the states of a quantum error-correcting
code can be ground states.  There is no inbetween.

\section{Discussion}
\label{sec:discussion}

%
%
We've developed several general results demonstrating that systems
exhibiting ground-state entanglement or correlation that is
``long-range'', in the sense of being between subsystems not directly
coupled, must necessarily have a small energy gap.  These results
suggest many interesting avenues for further investigation.

%
%
\textbf{Characterizing the physical properties responsible for the
  vanishing gap:} We have demonstrated several connections linking the
energy gap to long-range correlations and entanglement in the ground
state.  However, many of the connections we have identified only hold
for special (albeit still rather general) cases, rather than in the
most general case.  What are the physical properties responsible for
the vanishing of the gap in the most general case?

%
%
\textbf{Characterizing $R_G(\rho)$:} Our work has highlighted the
importance of understanding the set $R_G(\rho)$, defined to be the set
of all density matrices $\sigma$ with the property that
$\mbox{tr}_{\overline e}(\rho) = \mbox{tr}_{\overline e}(\sigma)$ for
all sets of systems, $e$, coupled by the coupling topology $G$.  In
physical terms, $R_G(\rho)$ contains all those density matrices
$\sigma$ which are energetically indistinguishable from $\rho$ for any
Hamiltonian respecting the coupling topology $G$.  Developing a good
mathematical and physical understanding of $R_G(\rho)$ is an extremely
challenging and interesting problem in quantum information science.
Promising preliminary work on this problem has been done
in~\cite{Linden02a,Linden02b}, but much remains to be done.

%
%
\textbf{The thermodynamic limit:} In the thermodynamic limit of a
large number of systems, the energy difference, $E_{\rm tot}$, between
the maximal and minimal energies in the system typically tends toward
infinity.  Recall that the results obtained in this paper typically
bound $\Delta E / E_{\rm tot}$ above by some measure of long-range
correlation, where $\Delta E$ is the energy gap.  Since $E_{\rm tot}$
tends to infinity in the thermodynamic limit, it follows that our
results do not give interesting information in this limit, except in
the case where we require exact ground states, i.e., $F = 1$.  It
would be extremely interesting to develop more powerful results
relating the gap to long-range correlations and entanglement in the
thermodynamic limit.

%
%
\textbf{Connection between the gap and the range of correlations:} We
have used ``long-range'' to mean entanglement or correlation between
parts of a system that are not directly coupled.  Of course, we expect
there will be substantial differences between a situation where two
subsystems are close, e.g., have perhaps a single spin mediating their
indirect interaction, and cases where the interaction is much more
indirect, e.g., the left- and right-hand ends of a linear chain, with
a large block of intermediate spins mediating the interaction between
the two ends.  We expect that the latter case will impose much more
stringent restrictions on the size of the gap than the former case.
Preliminary numerical investigations with the Heisenberg model bear
this out, and further investigations are currently underway.

%
%
In conclusion, we have used the techniques of quantum information
science to develop connections between the energy gap and long-range
correlations and entanglement in the ground states of many-body
quantum systems.  We believe that the techniques of quantum
information science will, more generally, be a powerful tool for
understanding and predicting the properties of complex quantum
systems.

\acknowledgments

We thank Dave Bacon, Andrew Doherty, Gerardo Ortiz, John Preskill,
Guifre Vidal and, especially, Nick Bonesteel, for interesting and
enjoyable discussions.


\end{document}